\documentclass[10pt,journal,a4paper,twoside]{IMEJ}
\usepackage{fdsymbol} % added

\usepackage[numbers, square, sort&compress]{natbib}

\usepackage[pdftex]{graphicx}

\usepackage{textcomp}

\usepackage{amsmath}
\usepackage{booktabs}
\usepackage{threeparttable}

\usepackage[caption=false,font=footnotesize]{subfig}

\usepackage{xcolor}
\usepackage{url}
\usepackage{balance}
\usepackage{comment}

\begin{document}

\renewcommand{\thepage}{\arabic{page}}% <-- ***REPLACE paper\_id WITH YOUR PAPER ID NUMBER***  

\bstctlcite{IEEEexample:BSTcontrol}

\title{Optimizing the Interplay Between the Chord-to-Radius Ratio, Camber and Pitch of Cross-Flow Turbine Blades}

\author{Ari Athair, Aidan Hunt, Han-Wen Chi, and
        Owen Williams% <-this % stops a space
\thanks{Preprint submitted to the 2025, 16$^{th}$ Annual European Wave and Tidal Energy Conference. The version that appeared in the conference preceding can be found at \protect\url{https://doi.org/10.36688/ewtec-2025-995}}%
\thanks{This work was supported in part by the U.S.A. DoD Grant N0002421D6400/N0002423F8719}% <-this % stops a space
\thanks{A. A. and O. W. are with the William E. Boeing Department of Aeronautics and Astronautics at the University of Washington, Author H. C. was previously as well, Steven's Way, Box 352400 Seattle, WA 98195 U.S.A. (e-mail: aristone@uw.edu, ojhw@uw.edu, michaelchi0608@gmail.com, respectively).}% <-this % stops a space
\thanks{Author A. H. is with the Department of Mechanical Engineering at the University of Washington, Steven's Way, Box 352600 Seattle, WA 98195 U.S.A. (e-mail: ahunt94@uw.edu)}
}% 

\markboth{}
{Athair \MakeLowercase{\textit{et al.}}: Optimizing the Interplay Between the c/R, Camber and Pitch of Cross-Flow Turbine Blades}

\maketitle

\begin{abstract}
This study examines the combined impact of the chord-to-radius ratio ($c/R$), blade camber, and preset pitch on cross-flow turbine performance. While prior research has examined individual effects of $c/R$ and pitch, this work focuses on their interaction with camber and the resulting aerodynamic behaviors. Central to the analysis is the concept of virtual camber and virtual incidence, which arise from the curved trajectory of the blades, and are quantified using a conformal transformation. Using two experimental datasets — one varying preset pitch, $c/R$, blade count, and Reynolds number, and the other examining geometric blade camber and pitch under identical conditions — this work demonstrates links between blade pitch and camber influences. We propose a potential pathway for reducing the design space by combining geometric variables and coupled virtual effects into two effective parameters: net camber and net incidence. These composite variables provide a more unified and descriptive framework for understanding turbine performance and design behavior, offering a pathway for more consistent optimization of turbine geometry.

\end{abstract}

% 3-5 Keywords
\begin{IMEJkeywords}
Cross-flow Turbines, Marine Renewable Energy, Turbine Blade Optimization, Virtual Camber, Flow Curvature Effects
\end{IMEJkeywords}

\section{Introduction}
\begin{figure} %
\centering
\includegraphics[width=86mm]{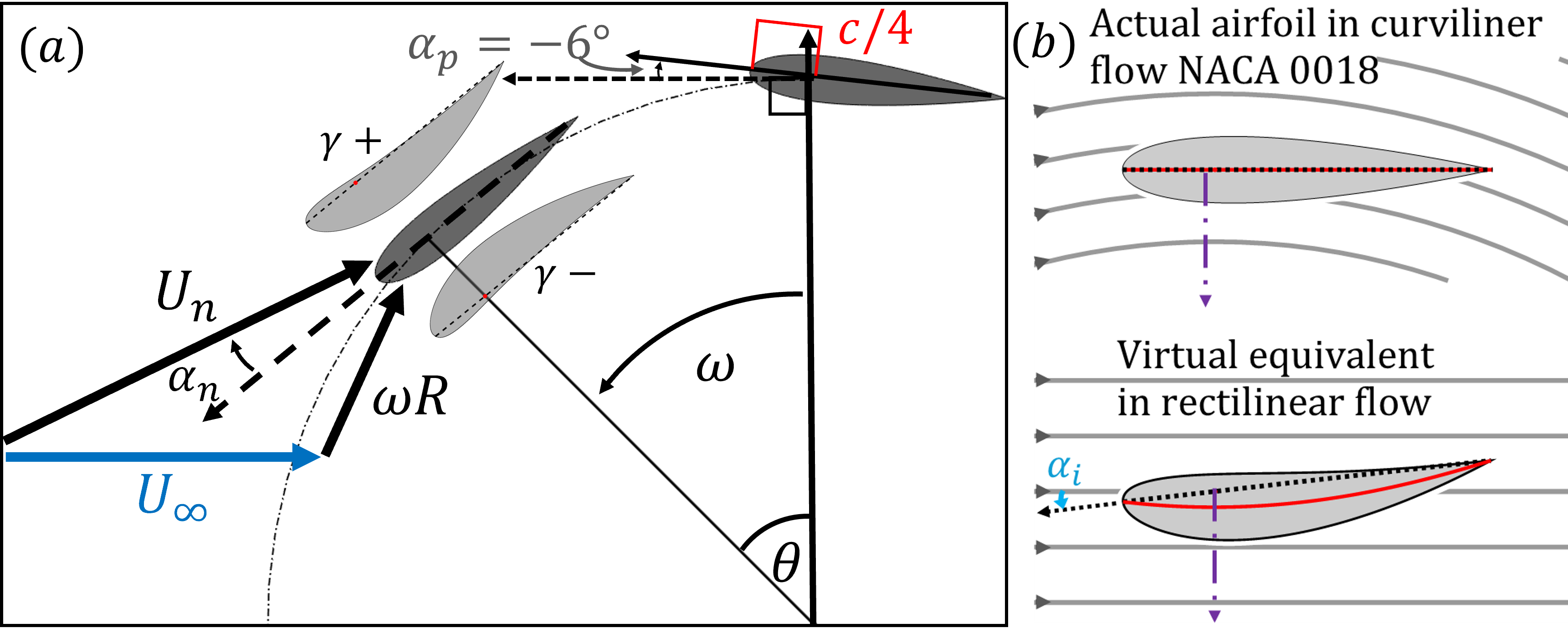}
\caption{(a) Cross section of turbine rotor to illustrate coordinate system, key variables and sign convention of cambered blades. Note that positive camber produces increased lift in the power-producing portion of the rotation by this convention. (b) Nominal representation of virtual camber and incidence, and how they arise from the curved blade path experienced by cross-flow turbines.}
\label{Fig:IntroCoord_VC}
\end{figure}

\IMEJPARstart{C}{ross-flow} turbines have significant potential for future renewable energy generation. Direction-insensitive operation, low rotation rates, confinement exploiting capabilities, and reduced environmental impacts make cross-flow turbines a desirable technology for marine energy applications \cite{Dabiri2011,Zhao2022,hunt2023BlockageEWTEC}. However, the development of an optimal cross-flow turbine design remains a significant challenge due to complex interactions between geometry and near-blade hydrodynamics, resulting in few generalized geometric design principles. While some studies have looked at the effects of the ratio of the blade chord and radius ($c/R$), blade preset pitch ($\alpha_p$), and blade camber ($\gamma$) individually (shown in Fig. \ref{Fig:IntroCoord_VC}a), these parameters are inherently interdependent and have not been considered collectively. In addition, the rotational motion of turbine blades introduces virtual camber and incidence effects (Fig. \ref{Fig:IntroCoord_VC}b) \cite{Migliore1980}, phenomena that are closely tied to $c/R$ but have not yet been integrated into a comprehensive design framework. The objective of this work is to analyze the individual and combined impacts of $c/R$, airfoil camber, and preset pitch on cross-flow turbine performance. By clarifying the connections between these geometric parameters and the resulting virtual effects, we aim to advance a more unified understanding of cross-flow turbine optimization and provide practical insights for future designs. 

\begin{figure*} %
\begin{center}
\includegraphics[width=168mm]{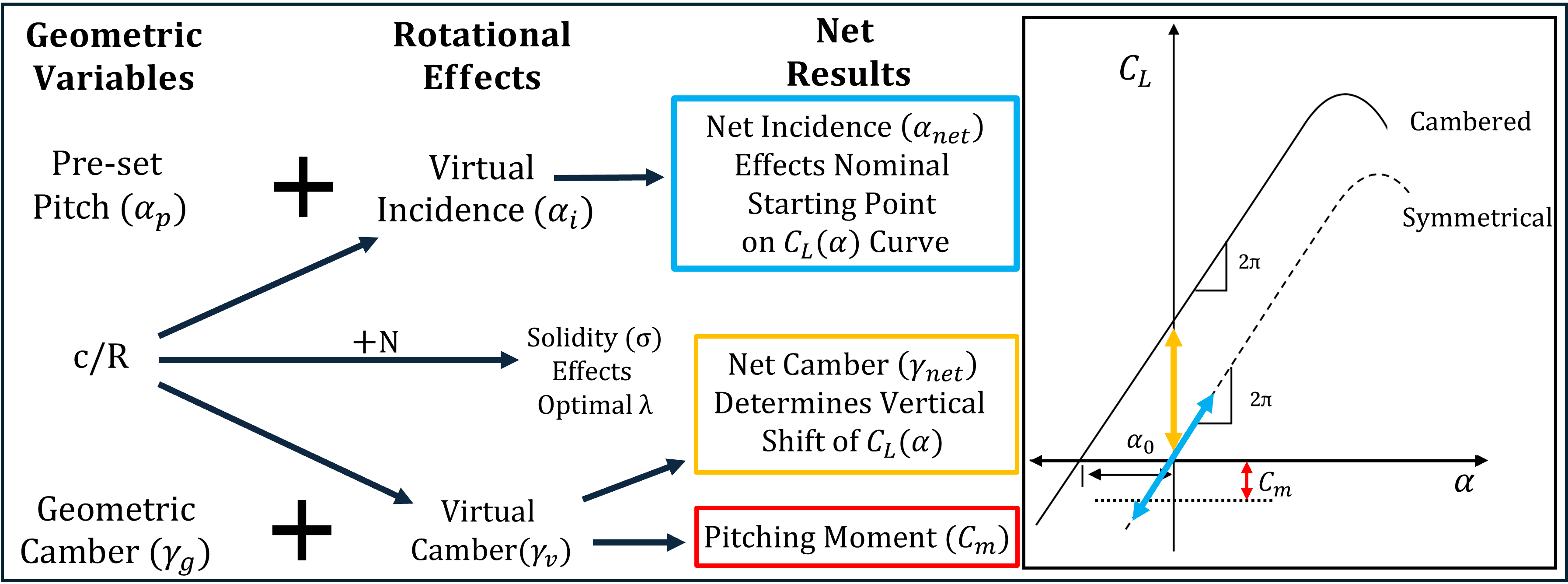}
\caption{Definition of net camber and incidence and their expected influence on blade lift and pitching moment from traditional non-rotating steady aerodynamic theory.}% 
\end{center}
\label{Fig:FlowDiagram}
\end{figure*}

In traditional, steady aerodynamics, blade pitch corresponds to a shift in angle-of-attack, resulting in movement of the nominal operating point of an airfoil diagonally along the $C_L-\alpha$ curve, shown in Fig. \ref{Fig:FlowDiagram} (blue). For a cross-flow turbine, changing preset pitch would therefore represent a change in blade lift at the start of its rotation, assuming dynamic effects were negligible. Conversely, steady aerodynamic theory suggests that airfoil camber produces a shift of the whole lift curve up and to the left (orange) and an increasingly negative pitching moment (red). Each change of these variables has a secondary effect on the range of angles of attack where the flow will remain attached, which is also important in cross-flow turbine performance. It is clear from this simplified model of incidence and camber effects that these parameters are linked by their effect on lift production and must be considered in conjunction to achieve a better understanding of aerodynamic behavior and an optimal performing design.

\begin{figure} %
\centering
\includegraphics[width=86mm]{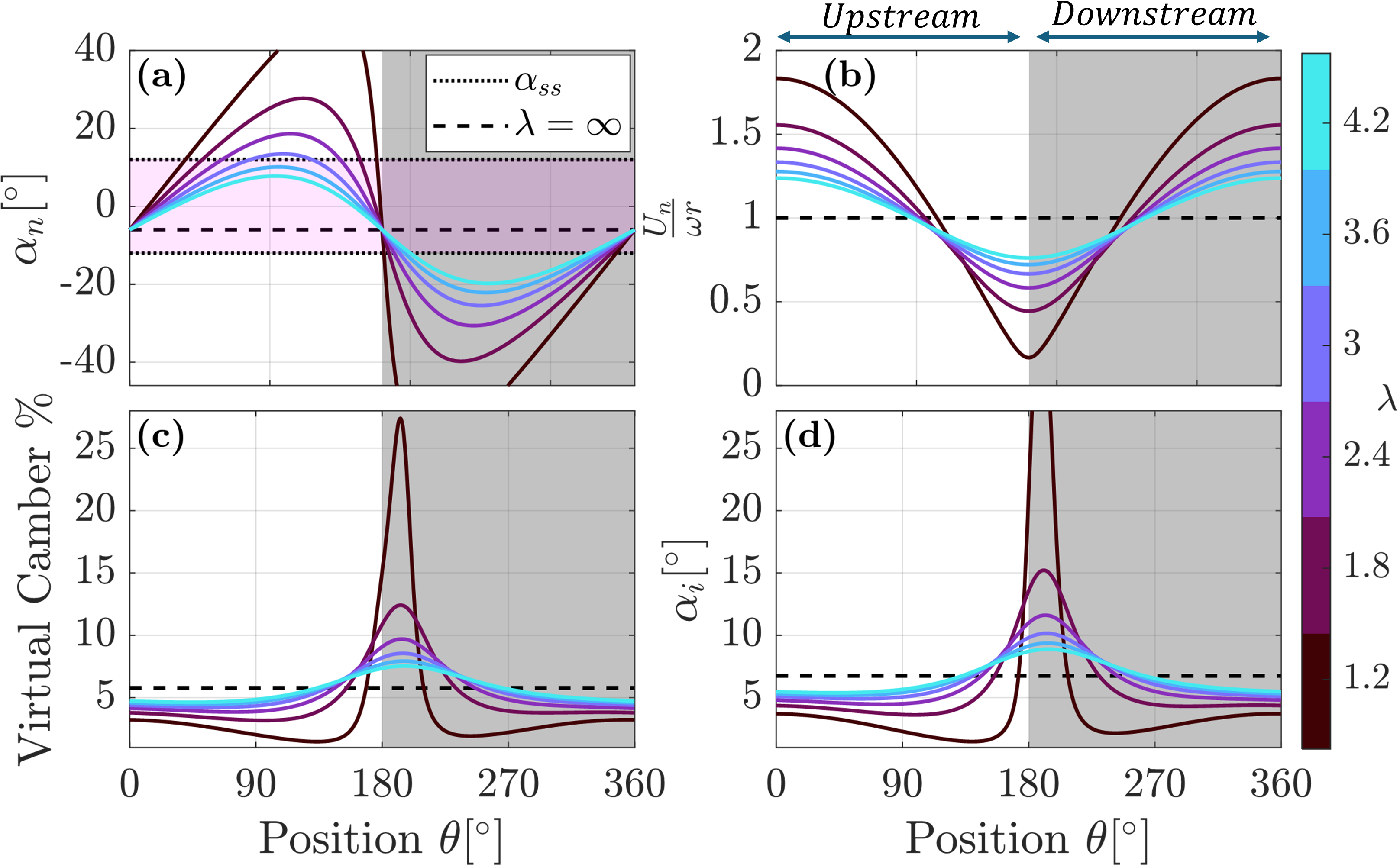}
\caption{(a) Nominal angle-of-attack, the dotted line and light pink shaded region correspond to the static stall angle ($\alpha_{ss}$) for a NACA 0018, $\pm$12$^\circ$ under similar Reynolds numbers to the cases explored \cite{timmer2008AoA,bianchini2016experimentalAOA}, (b) normalized blade relative velocity,(c) virtual camber and (d) virtual incidence ($\alpha_i$) as a function of position and $\lambda$ (lines). For a turbine of $c/R = 0.49$ and preset pitch $\alpha_p = -6^\circ$, $\lambda \rightarrow \infty$ horizontal black dashed lines. Grey shaded regions highlight the downstream region.}
\label{Fig:AoaUrelVCalphai_vs_TSR}
\end{figure}

However, making comparisons to traditional rectilinear airfoil theory and thin airfoil theory is a considerable oversimplification of the complexity present in cross-flow turbines. During turbine operation, the blades experience a wide range of angles of attack, nominally defined by 
\begin{equation}
  \alpha_n(\theta) = -tan^{-1}[\frac{sin(\theta)}{\lambda+cos(\theta)}]+\alpha_p  
  \label{eq:aoa}
\end{equation}%
and blade relative velocities, 
\begin{equation}
  U_n(\theta) = \sqrt{\lambda^2+2\lambda cos(\theta)+1} 
  \label{eq:Urel}
\end{equation}%
where $\theta$ is the blade position defined to be zero when the blade is pointing upstream (see Fig. \ref{Fig:IntroCoord_VC}a), and $\lambda = \omega R/U_\infty$ is the ratio of blade tangential velocity ($\omega R$) to the inflow velocity ($U_\infty$), known as the tip-speed ratio, which determines the near blade dynamics and overall performance. In both above equations, turbine induction has been neglected. Fig. \ref{Fig:AoaUrelVCalphai_vs_TSR}a-b shows curves of $\alpha_n$ and $U_n$ as a function of the tip-speed ratio. At low $\lambda$, the nominal angle-of-attack exceeds the static stall angle for rectilinear flow ($\pm 12^{\circ}$), which leads to strong unsteady and non-linear dynamic stall and flow reattachment process, which can be difficult to model. 

In the upstream region, $\theta = 0-180^{\circ}$, positive $\alpha_n$ is observed, lift dominates turbine loads and the majority of power production occurs. Conversely, in the downstream region, $\theta = 180-360^{\circ}$, the angle-of-attack becomes negative and performance depends on the ability to minimize power losses and improve flow recovery \cite{snortland2025downstream}. \citeauthor{snortland2025Force_arXiv} has shown that the pitching moment about the quarter-chord of the blade that arises from the virtual camber geometry, opposes power production and accelerates degradation of the downstream region at high $\lambda$. As $\lambda$ increases toward infinity (the limiting case of no inflow), nominal values approach a steady-state condition more closely approximating the classical aerodynamic arguments presented earlier. Induction (slowing of the flow around the turbine) leads the true angle-of-attack and relative velocity to converge towards the high $\lambda$ limit more quickly than nominal calculations would suggest, particularly in the downstream, where the momentum is reduced, lending some weight to the simplified expectations of Fig. \ref{Fig:FlowDiagram}. 

Despite the observation of flow curvature phenomena, its impact has often been neglected. In 1980, \citet{Migliore1980} explored the impact of a curvilinear flow on symmetrical blades, resulting in a transformation that maintains the local angle-of-attack along the chord. The result is a transformed blade which exhibits aerodynamic behavior akin to a 'virtual' cambered blade at a 'virtual incidence' ($\alpha_i)$ in rectilinear flow (Fig. \ref{Fig:IntroCoord_VC}b).

\citeauthor{Migliore1980}'s transformation utilizes inputs of $c/R$, $\alpha_p$, and blade mounting point, as well as $\lambda$ and $\theta$ to calculate the virtual equivalent foil and its variation throughout the cycle. The important outputs of the transformation to this work are the virtual camber, $\gamma_v$, and virtual incidence, $\alpha_i$. The relative curvature of the blade is parameterized by $c/R$, and therefore plays a dominant role in determining the magnitude of the virtual camber. Preset pitch and blade mounting point (which provides a datum for $\alpha_p$) also have secondary effects. For orientation, the virtual camber of a cross-flow turbine blade is positioned such that it would have an outward concavity away from the turbine axis throughout the cycle. We define this direction of curvature to be 'positive' as it enhances lift in the upstream portion of the cycle. Virtual incidence shows a stronger sensitivity to the mounting point and $c/R$ than the virtual camber, with preset pitch playing a lesser role. 

During turbine operation, a uniform inflow $(U_\infty)$ is superimposed on rotational flow, causing virtual camber and incidence to be functions of position and tip-speed ratio (see Fig. \ref{Fig:AoaUrelVCalphai_vs_TSR}c-d). Note that this position dependence is often simplified to result in a single measure of virtual camber or incidence for a given turbine geometry by assuming steady-state operation at an infinite tip-speed ratio (i.e no inflow). Convergence to this limit (horizontal dashed line) can be seen in Fig. \ref{Fig:AoaUrelVCalphai_vs_TSR}c-d.  This assumption is particularly appropriate for high $\lambda$ conditions and in the downstream region, where flow induction significantly reduces the effective inflow. Calculating at this limit provides position-independent values, useful in comparison with geometric design variables. Although \citeauthor{Migliore1980}'s method assumes a thin airfoil, neglects induction, and does not model stall, which are all present in a realistic cross-flow turbine, it remains a valuable tool for examining the effects of virtual geometry, and we adopt it in this paper to explore relationships between key parameters. 

To further explore the relationships between key parameters, we need to define two more variables. 
Net incidence, 
\begin{equation*}
    \alpha_{net}=\alpha_i + \alpha_p
\end{equation*}%
defined as the sum of the virtual incidence and the geometric preset pitch, 
and net camber, 
\begin{equation*}
    \gamma_{net}=\gamma_v + \gamma_g
\end{equation*}%
defined as the sum of the virtual camber and geometric camber, for a non-symmetrical blade. These parameters, and how they are expected to impact traditional aerodynamics, can be seen in Fig. \ref{Fig:FlowDiagram}. The definition of these net variables implicitly assumes that the impact of virtual and real geometry on the flow and performance is identical and that they can be linearly superimposed. This is a significant assumption that is yet to be demonstrated, but it is the simplest approach to combining these variables and one that appears reasonable based on results later in this paper.

\subsection{Review of relevant literature}

While cross-flow turbines have primarily employed symmetric blades, a wide range of camber magnitudes and directions (inward or outward) have been tested and suggested to be optimal in the literature, with little agreement \cite{Islam2007,letcher2017wind}. Many of these conflicting findings likely stem from testing a wide range of $c/R$ values and preset pitches without accounting for the resulting changes in virtual camber. In a broad sense, expectations from rectilinear flow apply, and positive camber increases upstream lift, while negative camber decreases downstream losses through drag and pitching moment reduction \cite{Danao2012,chi2024influence,athair2024APS}. However, the balance in performance upstream and downstream determines the time-average performance, and guidance for defining an optimal camber profile remains scant. \citet{Danao2012} and \citet{Danao2016MoreCambers} attempted to numerically optimize the blade shape through a range of negative camber and found the best performing version to be the profile that exactly canceled the virtual camber. In contrast, recent work at the University of Washington and University of Wisconsin-Madison has found some amount of positive net camber to be beneficial \cite{chi2024influence,athair2024APS}. \citet{chi2024influence} explored $\pm$2\% camber foil and compared the results to a symmetrical NACA 0018 for a range of $\alpha_p$. We make use of this dataset for the current analysis. Optimal camber has also been found to be tip-speed ratio dependent:  positive camber performs better below the peak $\lambda$ and negative camber superior above (for the tested profiles) \cite{athair2024APS}. These trends align with upstream and downstream mechanisms identified by \citet{snortland2025downstream}, where positive camber enhances upstream torque production to maximize performance at low rotation rates, while negative camber reduces downstream losses that determine power degradation at high $\lambda$.

The influence of preset pitch has received greater attention for symmetric foils. Recent experimental work by \citet{Hunt2024Geometry} evaluated the effects of preset pitch ($\alpha_p = -12^{\circ} - 0^{\circ}$), chord-to-radius ratio ($c/R = 0.25 - 0.74$) and blade count ($N = 1-4$) on NACA 0018 blades across four diameter-based Reynolds numbers between $\sim8\times$10$^4$ and $\sim8\times$10$^5$. A slight toe-out preset pitch (i.e., negative $\alpha_p$ measured from the quarter chord) consistently improved performance by reducing the upstream angle-of-attack, delaying stall. The optimal pitch was found to be strongly dependent on the $c/R$. As $c/R$ increased, a more negative preset pitch was found to be optimal. Furthermore, it has been hypothesized that by appropriately reducing the angle-of-attack, the preset pitch delays stall and reduces deep dynamic stall losses \cite{strom2015consequences}. \citet{Hunt2024Geometry} showed that a more negative preset pitch reduces the upstream performance peak but increases the downstream, and the optimal conditions were those that balanced the two regions.

 Notably, \citet{Hunt2024Geometry} also acknowledges and qualitatively discusses the increase in flow curvature effects with increased $c/R$ and compares it to limited prior works. Contrary to \citeauthor{Migliore1980,Mandal1994}, who suggested high $c/R$ (i.e. virtual camber) is detrimental, \citeauthor{Hunt2024Geometry} frequently found that the highest performance was achieved by the largest tested $c/R$, particularly at lower Reynolds numbers and blade counts. However, as the Reynolds number and blade count increased, lower $c/R$ blades became relatively more efficient. They also found that the solidity of the turbine, another non-dimensional parameter which changes with the change $c/R$ and number of blades, is linked to optimal tip-speed but is otherwise not a good predictor of performance as it can be achieved through multiple configurations \cite{Hunt2024Geometry}. 

 Some reduced-order modeling approaches, such as blade element momentum theory, require lift and drag polars from rectilinear flow. \citeauthor{Rainbird2015, Bianchini2018} have developed corrections to account for $c/R$ induced virtual camber and incidence. However, these still struggle to resolve performance predictions sufficiently accurately for geometric optimization purposes. While assessments have shown that virtual camber has a noticeable effect on performance and that it should be considered \cite{Migliore1980,Rainbird2015}, little research has explored how flow curvature analysis can deepen our understanding of performance outcomes.
 
 Prior work highlights an opportunity to apply a conformal transformation (such as \citeauthor{Migliore1980}) to expand quantitative analysis of $\alpha_p$ with $c/R$ and camber, linking performance more directly with both geometric and virtual geometries. In this paper, the interplay between the chord-to-radius ratio, camber, and preset pitch angle on cross-flow turbine performance is explored through cross-comparison of two experimental datasets: (1) performance data collected by \citet{Hunt2024Geometry} for turbines with symmetric NACA 0018 blades and various $c/R$, $\alpha_p$, and $N$, and (2) performance data collected by \citet{chi2024influence} for one- and two-bladed turbines with $\pm$2\% cambered NACA 0018 blades and various $\alpha_p$. Both datasets were obtained using the same experimental set-up and under similar flow conditions. In order to isolate the influence of geometric design parameters on turbine performance, each variable is kept constant in turn while varying others. Critically, the flow curvature (calculated using \citeauthor{Migliore1980} conformal transformation method) is accounted for when considering blade incidence and camber. This work aims to quantify the extent to which camber, $c/R$, and $\alpha_p$ are linked through flow curvature and whether their impact on performance is consistent across each variable.

\section{Experimental Methods}

\subsection{Turbine Test Set Up}

\begin{figure} %
\centering
\includegraphics[width=59mm]{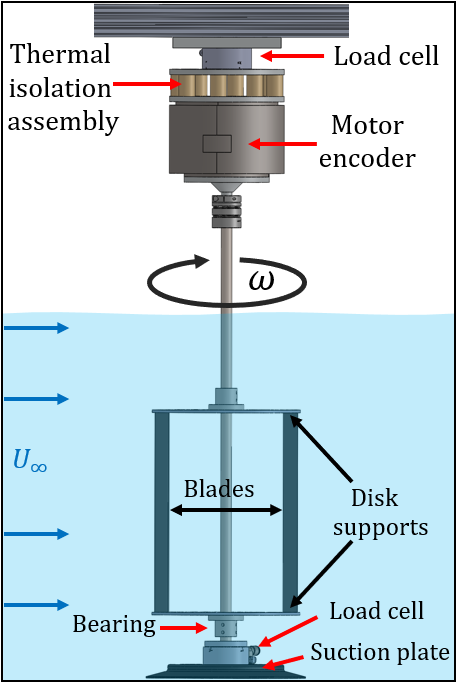}
\caption{Experimental test rig used in during all data collection.}
\label{Fig:ExpSetUp}
\end{figure}%

The experiments conducted by \citet{Hunt2024Geometry} and \citet{chi2024influence} were primarily performed in the Alice C. Tyler recirculating water flume at the University of Washington. The test section of the flume is 0.75 m wide and 4.88 m long. An integrated pool heater and chiller allow for control of the water temperature and thus, the viscosity of the freestream, and Reynolds number. The experimental test setup can be seen in Fig. \ref{Fig:ExpSetUp}. This work focuses on the cases with matching conditions between \citet{Hunt2024Geometry} and \citet{chi2024influence}. These cases had a 1.1 m/s free stream velocity, dynamic depth 0.48 m, temperature of 38.9$\pm0.1$ C$^\circ$, and a relatively low turbulence intensity of 2$-$4\%. The diameter-based Reynolds and channel depth, $H$, Froude numbers were held to $Re_{D} = D U_{\infty}/\nu = 2.7\times10^5$ and $Fr_h=U_\infty/\sqrt{g H}=0.51$, respectively, where $D=2R$ is the diameter of the turbine to the quarter chord. The blade height, $h$, and outermost diameter, $D_o$, were held constant at 23.4 cm and $\sim$17.2 cm, while the chord length, blade preset pitch, and blade profile were varied. The resulting blockage ratio was $\sim$11\% across all tests.

To economically evaluate a range of preset blade pitches, modular end plates with varying mounting patterns were employed, allowing blades of different chord lengths to be tested efficiently. Slight ($<$2\%) deviation in $R$ is present due to changing chord length and pitch due to the modular end plates; however, this is accounted for in performance calculation.
A servo motor mounted at the top of the center shaft regulates the rotation rate to a constant value. For each turbine configuration, a sweep of tip-speed ratios was performed. Data were logged using MATLAB at a sample frequency of 1 kHz, for further processing and analysis. Each test was run for a sufficient duration to ensure that the time-averaged performance converged and captured any cycle-to-cycle variability. During post-processing, data was trimmed to include an integer number of revolutions, and motor noise was filtered. Blade-level performance was extracted via phase-averaging as a function of blade position, while overall turbine performance was characterized by the time-averaged power coefficient ($\overline{C_P}$), where
\begin{equation}
    C_P = \frac{\tau \omega}{\frac{1}{2} \rho A U_\infty^3}
\end{equation}
defines the power coefficient as the ratio of mechanical power (torque times angular velocity), divided by the kinetic power in the flow passing through the projected area of the turbine. The outermost diameter defines the projected area of the turbine, $A=D_oh$.

For both datasets, supplemental experiments were conducted, with blades removed, under identical flow and control conditions to estimate losses. The results were then subtracted from the total turbine performance to isolate the blade level performance, following the superposition method identified by \citet{strom2018SuportStruct}.  Such that
\begin{equation}
    C_{P,blade} = C_{P,turbine}-C_{P,supports}.
    \label{eq:Cp_subtraction}
\end{equation}
A summary of all conditions tested in both studies is provided in Table \ref{tab:testconditions}. The turbine dimensions were consistent across both experiments; only the blade profiles changed, and one of the cases from \citet{Hunt2024Geometry} was run under matching non-dimensional parameters to \citet{chi2024influence}. Data from \citeauthor{Hunt2024Geometry} used in this work is openly available through the Dryad database \citet{Hunt2024Dryad}. Further detailed descriptions of the experimental setup and procedures are available in \citet{Hunt2024Geometry} and \citet{chi2024influence}.

\begin{table}
\centering
\caption{\label{tab:testconditions} Experimental conditions and geometry analyzed}
\begin{tabular}{l l  l  r r}
\toprule
&   \citet{Hunt2024Geometry}  & \citet{chi2024influence} 
\\\hline
$Re_{D} = D U_{\infty}/\nu$ [$\times10^5$] & 1.6 \& 2.7 & 2.7 \\
Blockage $\beta$ & 10.7$-$11.9 $\%$ & 11 $\%$ \\
Quarter chord radius $R$ [cm] & 8.15$-$8.24 & 8.15$-$8.24\\
Outermost Radius  $R_o$ [cm] & 8.23$-$8.98 & 8.6\\
Blade span $H$ [cm] & 23.4 & 23.4 \\
Blade Profile (NACA) & 0018 & $0018, \pm$2418 \\
Aspect Ratio (H/D) & 1.42 & 1.42 \\
Chord to Radius Ratio ($c/R$) & 0.25$-$0.74 & 0.49 \\
Preset pitch $\alpha_p$ & $-$12:2:0 &  $-$16:2:4\\
Number of Blades ($N$)  & 1$-$4 & 1$-$2 \\
\bottomrule
\end{tabular}
\end{table}%

\section{Results}

\begin{figure}
\centering
\includegraphics[width=86mm]{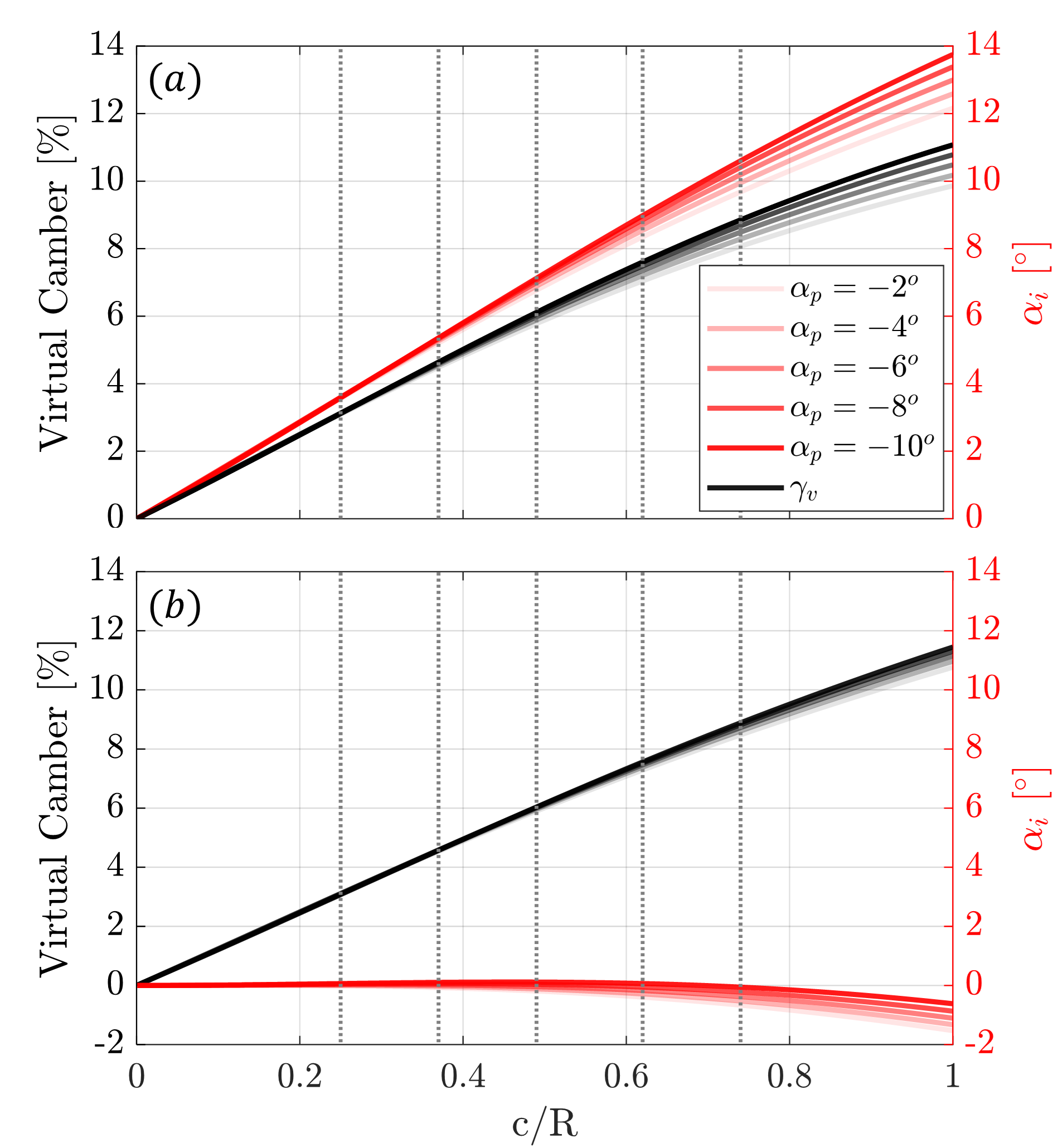}

\caption{(a) Effects of $c/R$ and $\alpha_p$ on virtual camber (black) and virtual incidence (red) at infinite $\lambda$ (no inflow) for a blade mounting point located at the airfoil quarter chord, and (b) mounting point at the half chord. Shading of the lines (red or black) corresponds to different preset pitch values. Vertical dotted lines correspond to the $c/R$ tested by \citet{Hunt2024Geometry}.}
\label{Fig:VC_CR}
\end{figure}

Utilizing the method of \citet{Migliore1980}, Fig. \ref{Fig:VC_CR} shows the effects of $c/R$ and $\alpha_p$ on virtual camber and virtual incidence at the limit of infinite tip speed (purely rotational flow). It is seen that as $c/R$ increases, both virtual incidence and virtual camber increase (corresponding to a toe-in pitching and concave-out camber, respectively). Both of which would be expected to increase power production in the upstream sweep, but be detrimental to performance in the downstream sweep \cite{Migliore1980}. While these flow curvature effects are predominantly influenced by $c/R$, changing the preset pitch angle has a small influence. Increasingly negative $\alpha_p$ (i.e., more toe-out pitch) resulting in higher virtual camber and incidence. This effect grows with $c/R$ and is negligible for short chord lengths.

As an aside, some turbines mount blades near the half-chord location and the change in reference impacts the calculated flow curvature effects (Fig. \ref{Fig:VC_CR}b). By transitioning from the quarter-chord to the mid-chord with a fixed geometry, the radius to the mounting point changes slightly (depending on the pitch and chord length), and the blade preset pitch measured at the mid-chord becomes more negative. Changing the reference point causes the calculated virtual camber to become nearly independent of preset pitch, and virtual incidence is zero for all but the highest $c/R$ values. We mention this difference to note that care should be taken during evaluation of virtual geometry. However,  most cross-flow turbines mount the blades at the quarter-chord and so all remaining analysis maintain this convention.

\begin{figure}
\centering
\includegraphics[width=86mm]{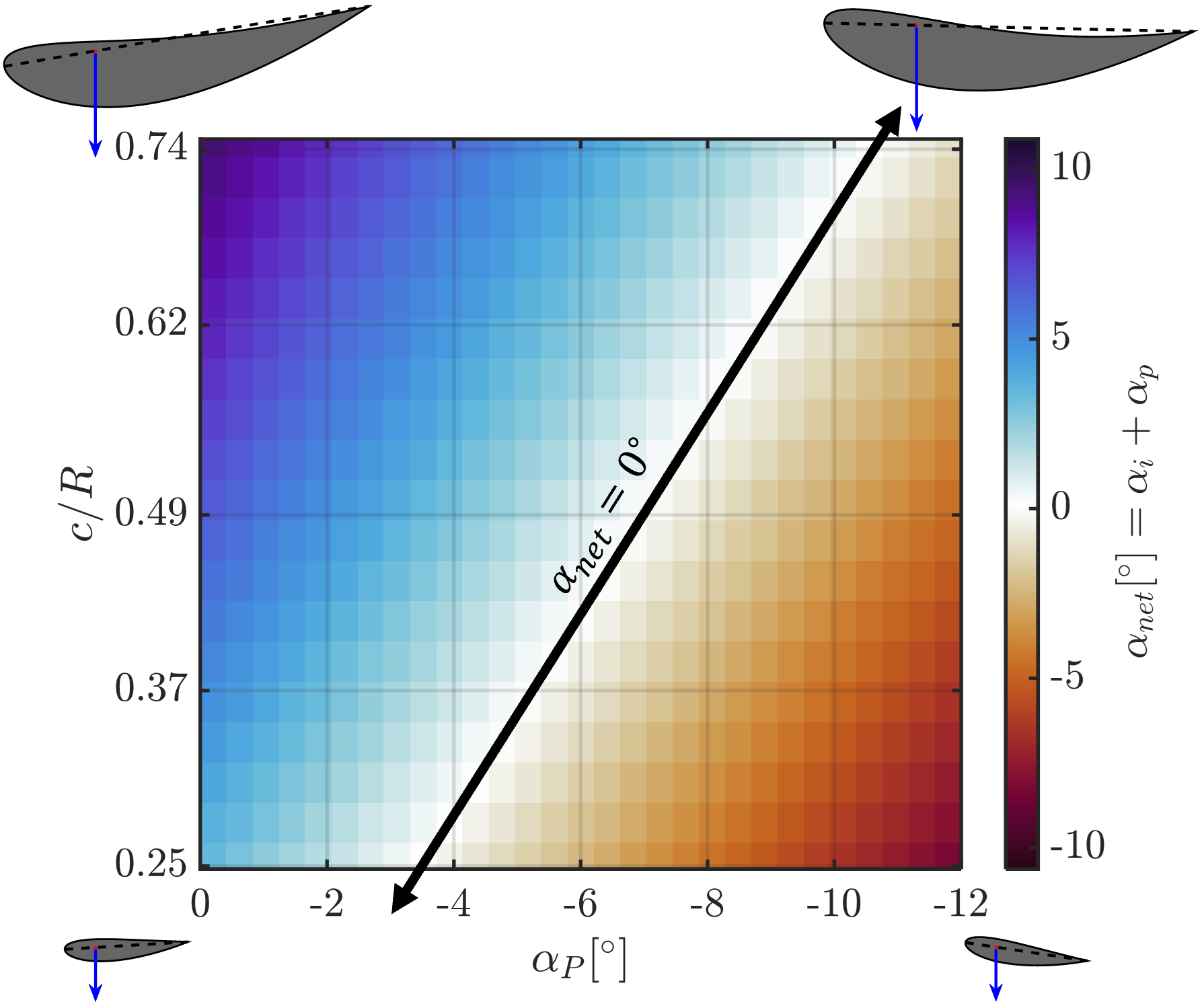}
\caption{Effects of $c/R$ and $\alpha_p$ on net incidence ($\alpha_i$). Calculated using the method of \citet{Migliore1980} assuming $\lambda = \infty$, and a blade mounting point at $c/4$. Blade shapes correspond to net geometry (virtual + geometric) present at each corner of the explored $c/R$ versus $\alpha_p$ space. The large black arrow indicates the line of 0$^\circ$ net incidence.}
\label{Fig:NetIncidence DesignSpace}
\end{figure}

Since both virtual camber and virtual incidence vary with $c/R$, isolating either effect requires attendant adjustments to $\alpha_p$ to hold the net incidence constant. By using the method of \citet{Migliore1980} to calculate the net incidence for the two-dimensional parameter space explored in \citet{Hunt2024Geometry}, a better understanding of the net geometry variation throughout the design space can be obtained. Fig. \ref{Fig:NetIncidence DesignSpace} shows the dramatic range of net geometries explored through combinations of $\alpha_p$ and $c/R$ tested by \citet{Hunt2024Geometry}. In order to decouple the net effects of geometry, it is necessary to calculate the effects of flow curvature from the $c/R$, and explore results along parameter iso-lines. Using Fig. \ref{Fig:NetIncidence DesignSpace} as a guide and choosing combinations of $c/R$ and $\alpha_p$ along, or parallel to, the black, diagonal arrow (constant $\alpha_i = 0^\circ$) effects of net incidence are fixed, and it becomes possible to isolate the impacts of the net camber at any point along the line. Interestingly, the diagonal line of constant net incidence closely follows trends in optimal $\alpha_p$ for the given $c/R$ configuration observed in Fig. 10 from \citet{Hunt2024Geometry}, suggesting that optimum preset pitch is largely determined by the corresponding virtual incidence at a given $c/R$. 

Holding net incidence fixed allows isolation of the effect of virtual camber due to increasing $c/R$, assuming dynamic effects and phase variation in $\alpha_n$ are negligible. Similarly, we assume net camber captures the full effect of geometric and virtual camber in one parameter. While it is a significant assumption that the simple summation of virtual and geometric factors best represents their combined influence on turbine performance, this is the simplest initial option and one that seems reasonable based on historic data and results further in this paper.

\begin{figure}[!t]
\centering
 \includegraphics[width=86mm]{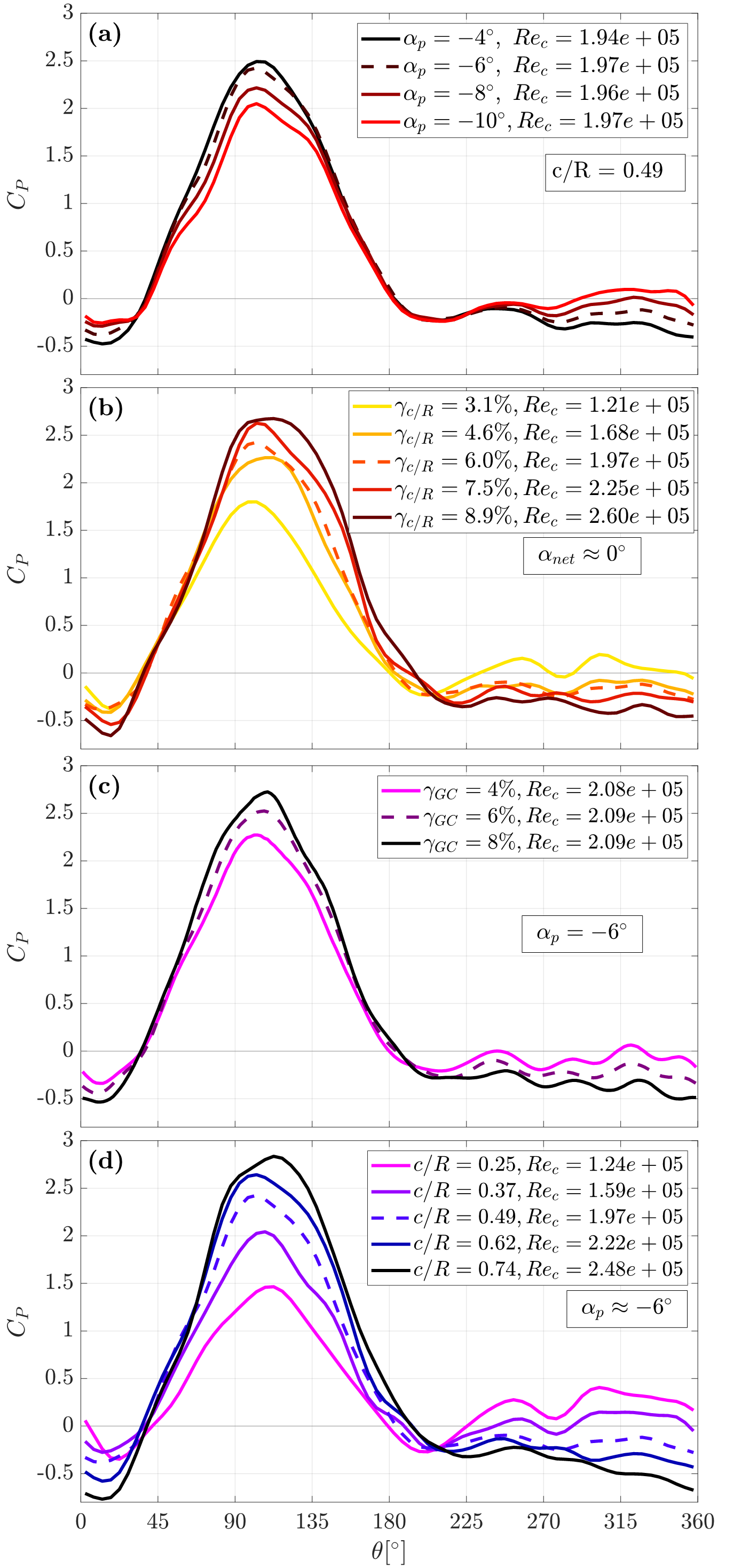}
\caption{Single-bladed phase-averaged performance for various turbine geometries at $Re_D = 2.7\times10^5$ and optimal tip-speed ratio as a function of changes to pitch and camber parameters. (a) Changes in blade pitch for fixed $c/R = 0.49$. (b) Effects of net camber through changes in $c/R$ $(\gamma_{c/R})$, with $\alpha_p$ varied to maintain approximately constant $\pm1^\circ$ net incidence. (c) Effects of net camber via changes in blade geometric camber $(\gamma_{GC})$ for fixed, $c/R = 0.49$, and $\alpha_p=-6^o$. (d) Performance when keeping $\alpha_p$=$-6^o$ fixed and varying $c/R$, resulting in changes to both net incidence and virtual camber, causing a compounded influence on the results. Each line corresponds to the optimal $\lambda$ for the given configuration, and the dashed line corresponds to consistent geometry across all four subplots. (a-b \& d) data are from \citet{Hunt2024Geometry} while (c) data are from \cite{chi2024influence}.}
\label{Fig:PhaseAvg}
\end{figure}

Fig. \ref{Fig:PhaseAvg} demonstrates the manner in which preset pitch, virtual camber, and geometric camber all cause similar trends in performance. Each curve corresponds to the optimal tip-speed-ratio for the tested design. Across all profiles, the optimal tip-speed-ratio ranged from 2.3$-$3.6, with the highest for low $c/R$, and the lowest for high $c/R$. Minimal difference in optimal tip-speed-ratio is seen across the geometric cambered cases. Increasing toe-out preset pitch (more negative $\alpha_p$) at constant $c/R$ causes lower performance in the upstream and higher downstream (\ref{Fig:PhaseAvg}a). This is attributed to a reduction in the nominal angle-of-attack during the upstream sweep, but an increase in the nominal angle-of-attack in the downstream sweep. Increasing net camber has a similar effect on the phase-averaged data through virtual or geometric camber (Fig. \ref{Fig:PhaseAvg}b,c), with higher camber increasing upstream power production for positive angles of attack and reducing it downstream, where negative angles of attack are seen. Solely changing $c/R$ while keeping $\alpha_p$ fixed (Fig. \ref{Fig:PhaseAvg}d) results in a larger change in peak phase-averaged performance compared to the other figures. This indicates that the combined increase in incidence and virtual camber are having a cumulative influence, as if the influences of Fig. \ref{Fig:PhaseAvg}a and \ref{Fig:PhaseAvg}c were being added together (i.e.increasing both camber and pitch in tandem). Across all phase-averaged cases, augmentations in the upstream are offset by reductions in performance in the downstream portion of the rotation, or vice versa. More positive camber or pitch results in increased upstream peak performance, but is paired with a drop in performance downstream. The key takeaway from the analysis of this unique combination of data is that all of these parameters exhibit similar effects on performance. Similarly, the magnitude of change produced by each of these variables is also comparable. As a result, they should be considered in conjunction during the design and optimization of turbine geometry; something rarely done at present. 

It should be noted that while the diameter-based Reynolds number and blockage have been held constant, the influences of Reynolds number and blockage on these datasets must still be considered, as the blade local Reynolds number changes with chord length and turbine thrust changes turbine induction. As a result, we must explore whether the changes observed in Fig. \ref{Fig:PhaseAvg} might be the result of Reynolds number or blockage effects, which could not be wholly removed. To explore these influences, the Reynolds number, $Re_c$, is also calculated and included in Fig. \ref{Fig:PhaseAvg}, based on the blade chord length and nominal blade-relative velocity. The datasets of the current work are not in a Reynolds number independent regime, according to prior works, which found diameter-based Reynolds number independence to occur above a value of $\approx10^6$ or chord-based Reynolds number of 1.5-2$\times10^5$
\cite{miller2021solidity,miller2018vertical,bachant2016effects,Ross2022}. In the range of Reynolds numbers observed in this study, cases with higher $Re_{c, rel}$ is expected to maintain flow attachment longer, and increase performance. Changes in $c/R$ and $\lambda$ have opposing effects on $Re_c$: increasing $c/R$ increases $c$, while decreasing optimal $\lambda$, which decreases the mean relative velocity. For the current datasets, the effects of increasing $c$ dominate and cause $Re_{c, rel}$ to grow. 

While Reynolds number influences remain in the comparison of Fig. \ref{Fig:PhaseAvg}, there are multiple reasons to believe that they have a secondary influence on the observed trends. First, when comparing the geometric camber cases (Fig. \ref{Fig:PhaseAvg}c) where $Re_c$ is approximately constant, we still see strong matching to curves of similar virtual equivalence from changes in $c/R$ (Fig. \ref{Fig:PhaseAvg}b), suggesting that changes in $Re_c$ are having a small impact relative to flow curvature effects for $Re_c >1 \times 10^5$. In addition, examination of the broader  dataset from \citet{Hunt2024Dryad}, suggests that the magnitude of change in $Re_c$ between different $c/R$ cases is smaller than would explain the amount of change in performance for $Re_c = 1.6-2.7 \times 10^5$. For equivalent physical geometries tested at $Re_c=2.60 \times 10^5$ and $Re_c=1.45 \times 10^5$, the difference in peak phase-average $C_P(\theta)$ at optimal $\lambda$ is observed to be 0.26 by \citet{Hunt2024Dryad}. In contrast, for a fixed diameter-based Reynolds number but varying virtual geometries the change in peak $C_P(\theta)$ is 0.88, shown in Fig. \ref{Fig:PhaseAvg}b. The range of tested $c/R$ corresponded to associated chord-based Reynolds numbers of $Re_c=2.60 \times 10^5$ and $Re_c=1.21 \times 10^5$. This result suggests that the changes in virtual geometry lead to a 3.3 times greater change in peak performance than might be expected from the change in Reynolds number alone.

Blockage effects should also be considered when analyzing and comparing performance across these experiments. Prior work has shown blockage to have minimal impact below a threshold of 10\% \cite{goude2014simulations}. The data in this study were collected at a slightly higher blockage ratio of $\sim$11\%, but its effects are expected to be minor compared to the examined design factors such as pitch, flow curvature, and blade camber. Although blockage was held constant across experiments, performance augmentation also depends on the turbine’s thrust coefficient. Turbines with higher thrust—for example, those with larger $c/R$—experience greater performance gains. Single-bladed turbines, the primary focus of this work, obstruct less flow and are therefore more robust to blockage effects than multi-bladed designs \cite{hunt2025HighBlockageArray}. While the influences of blockage cannot be entirely eliminated from the comparisons of the current study, we believe them to be less significant than those due to geometry.

\begin{figure*}
\centering
\includegraphics[width=130mm]{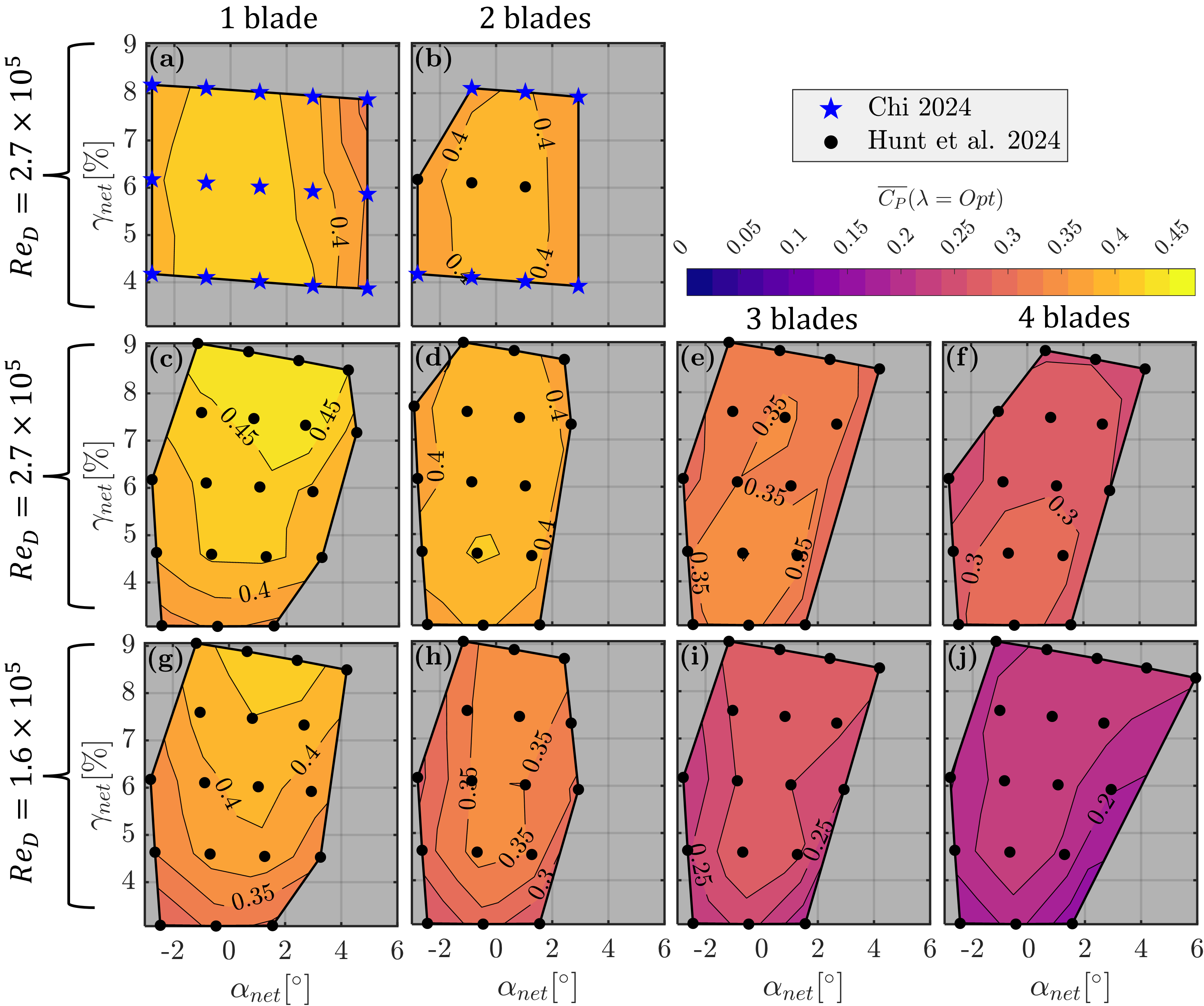}
\caption{Heat maps of blade level performance as a function of net camber ($\gamma_{net}$) and net incidence ($\alpha_{net}$) from \citet{Hunt2024Geometry} and \citet{chi2024influence}. Columns correspond to the number of blades. The top row corresponds to cases where $c/R = 0.49$ is kept constant and the net camber is changed by using blades with geometric camber from \citet{chi2024influence} points denoted as $\color{blue}{\medblackstar}$. Data from \citet{Hunt2024Geometry}, indicated by $\color{black}{\medblackcircle}$, whose experiments utilized symmetric blades and varied $c/R$.  For these rows, the change in net camber is therefore due solely to a change in $c/R$.}
\label{Fig:PerfHeatMap_vs_flowcurvature}
\end{figure*}

 We note that the diagonal line of net zero incidence is observed in Fig.  \ref{Fig:NetIncidence DesignSpace} corresponds closely to a band of optimum performance for a given $c/R$ observed by \citet{Hunt2024Geometry}. To explore this further, the data of \citet{Hunt2024Geometry} are replotted in Fig. \ref{Fig:PerfHeatMap_vs_flowcurvature} and compared to the data of \citet{chi2024influence}, by replacing preset pitch and $c/R$ with net incidence and camber. Across the combinations of $c/R$, $\alpha_p$, $N$, and $Re_D$ tested by \citeauthor{Hunt2024Geometry}, it is seen that the optimal performance at a given $\gamma_net$ (from either $c/R$ or blade camber)  occurs at a relatively constant net incidence of $0^\circ$ to $+1^\circ$.  As a result, the virtual incidence seems to be the main driver of optimal $\alpha_p$ at a given $c/R$ or camber. This result provides a theoretical basis for the selection of preset pitch, which was previously optimized empirically. 

Although it may be intuitive that virtual and geometric pitch seem the most coupled, this result is surprising when the marked difference in virtual camber present across the tested parameter space is considered. Despite the expected change in fluid dynamic performance of the blades due to increased lift and opposing pitching moment from camber, optimum pitch appears to be independent of geometric camber (shown by the vertically shaped contours shown in Fig. \ref{Fig:PerfHeatMap_vs_flowcurvature}a-b, where changes in net camber are solely the result of changes in blade geometric camber for fixed $c/R$). The fact that net camber, either virtual from $c/R$ or geometric camber, does not seem to make a notable impact on optimum pitch, suggests an avenue for further study.

While virtual camber derived from $c/R$ does not appear to impact the optimal pitch as a function of $c/R$, virtual camber has an impact on the optimum $C_P$. However, for the one-bladed data, we observe slightly different behaviors when manipulating net camber through changes in $c/R$ and through geometric camber. In Fig. \ref{Fig:PerfHeatMap_vs_flowcurvature}a, contours are primarily vertical and there is minimal difference in performance with changes in net camber. While in Fig. \ref{Fig:PerfHeatMap_vs_flowcurvature}c, as net camber increases due to larger $c/R$, the $C_P$ is also seen to increase. It is hard to say if this difference is due to changes in the manner in which geometric and virtual camber influence performance or if they are related to changes in $Re_c$ or flow induction. Changes in chord-based Reynolds number present in those datasets varying $c/R$, are not present in the geometric camber data. In addition, larger values of $c/R$ accentuate blockage/channel-turbine interactions. It is therefore not possible to determine if virtual and geometric camber behave differently using this dataset. 

The two-bladed data in contrast (Fig. \ref{Fig:PerfHeatMap_vs_flowcurvature}b,e), exhibit very similar vertical contours for both $c/R$ and geometric camber changes to net camber. And further extending to higher numbers of blades does not seem to alter trends in net incidence and camber notably. The sensitivity of $C_P$ to changes in $c/R$-induced virtual camber is also found to be highest for a single-bladed turbine and lessened for higher blade count.

\section{Discussion and Conclusions}

This work highlights the critical importance of incorporating virtual camber and incidence into the design and analysis of cross-flow turbine blades. By framing performance trends in terms of combined metrics of net camber and net incidence, we gain a more comprehensive understanding of how $c/R$, $\alpha_p$, and camber influence turbine behavior. This approach provides a powerful lens for interpreting results across a wide range of conditions and geometries.

A key outcome of this analysis is the identification of net incidence as a primary predictor of optimal blade pitch for a given chord-to-radius ratio ($c/R$), independent of the blade’s geometric camber. It is perhaps surprising that net incidence appears to predict optimum blade ($\alpha_p$), given the simultaneous change in net camber which occurs across a range of $c/R$. Furthermore, our net camber and incidence metrics are based on infinite $\lambda$, and a transformation that assumes purely rotational flow and thin foils. And yet, calculating net incidence still provides a clear pathway to calculate optimal $\alpha_p$. While there are other parameters that could be expected to influence the optimum $\alpha_p$, they do not seem to matter here.

The data suggest that while geometric camber and $c/R$ both have a comparable phase-averaged effects when examined through net camber,  this metric cannot be fully optimized with the current analysis. Net camber is shown to affect performance by changing the weighting of the phase-averaged upstream and downstream performance, with more positive (more concave-out) camber enhancing power production, while lower net camber (less concave-out) sees better performance downstream. However, it appears the ideal $c/R$ may be due to a combined effect of geometric camber, virtual camber, and changes in pitching moment (which is detrimental to performance), which is poorly predicted at this point. Overall performance will be determined by the balance of these effects in the upstream and downstream regions, which will require future exploration. 

Depending on whether net camber is achieved through changes in $c/R$ or geometry, it is unclear if blades should be optimized for higher net camber or if the optimum $C_P$ is fairly insensitive to net camber, since blade-relative Reynolds numbers and channel-turbine interactions/induction could not be held constant across all data. Despite this, net positive camber appears optimal overall, and the sensitivity to $c/R$-induced virtual camber appears greatest at for a single-bladed turbine.

This work also highlights a pitfall associated with mounting blades at the traditional quarter-chord location. In rectilinear aerodynamics under non-stall conditions, the aerodynamic center and thus center of pressure is located at the quarter chord. Because of this, the pitching moment is expected to be negligible about that point for symmetric foils, and so only lift and drag are relevant to aerodynamic analysis. However, in cross-flow turbines with virtual camber, by designating a quarter-chord mounting point, a more significant change in angle-of-attack is introduced along the length of the blade. By changing the typical blade mounting point coordinate system to the half-chord, we show that virtual incidence can be largely eliminated from the variables necessary in the analysis of net geometry. This indicates that the virtual incidence effectively behaves as a correction factor for the preset pitch if the mounting point is offset from the mid-chord of the blade. Moving forward, care should be taken in the documentation of the reference mounting point, particularly for blade-level analysis of data, as the net geometry can have a large impact. 

As virtual incidence appears to determine the optimal preset pitch, and net camber may be used to provide insights into the optimal camber of turbine blades. It opens the question of whether these can be combined into a single metric. An angle often associated the $C_L$-$\alpha$ curve of a cambered foil is the zero-lift angle-of-attack or $\alpha_0$, the angle at which no lift is produced for a particular airfoil. Thin airfoil theory provides a means of approximating $\alpha_0$ with only the camber line as input. By doing so, net camber could be reduced to an angular value and combined with net incidence. However, at this point, it is unclear if this is useful, as pitch and camber have a distinctly different effect in traditional aerodynamics. Further exploration and optimization of the consideration of net camber and net incidence should be evaluated separately. This does allow reducing the parameter space to two variables that describe the combined performance of the original three variables ($\alpha_p$, $c/R$, and geometric camber), which are coupled due to flow curvature.

In future studies, data captured for a range of $c/R$, geometric camber and $\alpha_p$, each modified in conjunction, would allow testing of the effect of net conditions on performance. Exploration of the changing ratio of geometric and virtual camber, while maintaining net camber, could shed light on any nonlinear interactions between virtual and geometric parameters not captured in this work, as well as evaluate the secondary influences of chord-based Reynolds number and blockage/channel-turbine interactions. Finally, while this work focuses on mean and peak efficiency trends due to changes in net geometry, these parameters are also expected to alter peak-to-average power and loading ratios, torque variability, and structural loads, which will affect generator efficiency and support structure costs. Exploration of these system-wide impacts on design would be important for future avenues of exploration.

%\vfill
%\newpage

\section*{Acknowledgement}
The authors would like to gratefully acknowledge
Abigale Snortland, Ph.D. of Pacific Northwest National Lab, Caelan Consing and Jennifer A. Franck, Ph.D. of the University of Wisconsin-Madison, and Brian Polagye, Ph.D. of the University of Washington, for in-depth discussions on Reynolds number and blockage effects in conjunction with the effects of camber, pitch, and flow behavior throughout the rotation of cross-flow turbines.

\ifCLASSOPTIONcaptionsoff
  \newpage
\fi

\bibliographystyle{IEEEtranN}

% balance columns on final page
\balance

% argument is your BibTeX string definitions (if any) and bibliography database(s)
%{\footnotesize
\bibliography{example}

%%% THE END %%%
\end{document}